\documentclass[%
 reprint,
superscriptaddress,
nobibnotes,
 amsmath,amssymb,
 aps,
prl,
citeautoscript,
floatfix,
showkeys
]{revtex4-2}

\usepackage[T1]{fontenc} 
\usepackage[]{graphicx}
\usepackage[utf8]{inputenc}
\usepackage{blindtext}
\usepackage{lmodern}%
\usepackage{hyperref}%
\usepackage{xcolor}%
\usepackage{placeins}
\usepackage{amsmath,amssymb}
\usepackage{bm}
\usepackage{tabularx}
\usepackage{booktabs}
\usepackage{notes2bib}
\usepackage[normalem]{ulem}
\usepackage[normalem]{ulem}
\usepackage{placeins}
\usepackage{amssymb}

\newcommand{\cm}{cm\ensuremath{^{-1}}}

\newcommand{\beq}{\begin{equation}\begin{aligned}}
\newcommand{\eeq}{\end{aligned}\end{equation}}

\newcommand{\nips}{NiPS\ensuremath{_3}}
\newcommand{\CrPlus}{Cr\ensuremath{^{3+}}}
\newcommand{\QuTtwo}{\ensuremath{^4}T\ensuremath{_2}}
\newcommand{\HaTtwo}{\ensuremath{^2}T\ensuremath{_2}}
\newcommand{\HaTone}{\ensuremath{^2}T\ensuremath{_1}}
\newcommand{\HaE}{\ensuremath{^2}E}
\newcommand{\QuAtwo}{\ensuremath{^4}A\ensuremath{_2}}

\newcommand{\CrIThree}{CrI\textsubscript{3}}

\newcommand{\CPS}{CrPS$_4$}

\newcommand{\Tn}{\ensuremath{T_{\rm N}}}

\newcommand{\neel}{\emph{Néel}}
\newcommand{\vdW}{\emph{van der Waals}}

\newcommand{\figref}[2]{\ref{#1}\textsf{#2}}

\newcommand{\ie}{\emph{i.e.},}

\newcommand{\dqmp}{Department of Quantum Matter Physics, University of Geneva, 24 Quai Ernest Ansermet, CH-1211 Geneva, Switzerland}
\newcommand{\gap}{Department of Applied Physics, University of Geneva, 24 Quai Ernest Ansermet, CH-1211 Geneva, Switzerland}
\newcommand{\nasu}{Advanced Materials Nonlinear Optical Diagnostics lab, Institute of Physics, NAS of Ukraine, 46 Nauky pr., 03028, Kyiv, Ukraine}

\definecolor{linkcol}{rgb}{0,0,0.4}
\definecolor{citecol}{rgb}{0.5,0,0}
\definecolor{harvardcrimson}{rgb}{0.79, 0.0, 0.09}
\definecolor{lava}{rgb}{0.81, 0.06, 0.13}

\hypersetup{
	bookmarksopen=true,
	pdftitle="Brightened Optical Transition as Indicator of Multiferroicity in a Layered Antiferromagnet",
	pdfauthor="et al",
	pdfsubject="", 
	pdfstartview={FitH}, 
	pdfmenubar=true, 
	pdfhighlight=/O, 
	colorlinks=true, 
	pdfpagemode=UseNone, 
	pdfpagelayout=SinglePage, 
	pdffitwindow=true, 
	linkcolor=harvardcrimson, 
	citecolor=harvardcrimson, 
	urlcolor=lava
}

	\begin{abstract}
    Two-dimensional van der Waals magnets show strong interconnection between their electrical, magnetic, and structural properties. Here we reveal the emergence of a luminescent transition upon crossing the \neel\ transition temperature of \CPS, a layered antiferromagnetic semiconductor. This luminescent transition occurs above the lowest absorption level. We attribute the optical transitions to excited states of the t$_{\rm 2g}$ orbitals of the \CrPlus\ ions, which are influenced by the distortion of the octahedral crystal field. Specifically, we find at the crossing of the \neel\ temperature changes the distortion from an anti-polar to polar arrangement, thereby not only activating an additional luminescent pathway but also inducing a significant in-plane static dipole moment detected by a marked enhancement in the intensity of the second harmonic generation. Our findings suggest the presence of a multiferroic state in \CPS\ below the \neel\ temperature.
    \end{abstract}

\begin{document}
	
	
	\title{Brightened Optical Transition as Indicator of Multiferroicity\\ in a Layered Antiferromagnet}


    \author{Volodymyr Multian}
    \affiliation{\dqmp}
    \affiliation{\gap}
    \affiliation{\nasu}
    \author{Fan Wu}     
    \affiliation{\dqmp}
    \affiliation{\gap}
    \author{Dirk van der Marel} 
    \affiliation{\dqmp} 
    \author{Nicolas~Ubrig}
    \affiliation{\dqmp}
	\email{nicolas.ubrig@unige.ch}
    \author{Jérémie Teyssier}
    \affiliation{\dqmp}

 	\date{\today}

	\maketitle

The discovery of magnetism in layered two-dimensional materials offers an unprecedented platform to tailor interactions between spin and charges, potentially leading to the discovery of multiferroic states~\cite{burch_magnetism_2018,gibertini_magnetic_2019,huang_emergent_2020}. The magnetic state of these materials can be modified in the same device by electric field~\cite{jiang_electric-field_2018,huang_electrical_2018}, doping~\cite{jiang_controlling_2018,wu_gate-controlled_2023}, or ultrafast optical light pulses~\cite{belvin_exciton-driven_2021,zhang_all-optical_2022,daabrowski_all-optical_2022,matthiesen_controlling_2023}, rather than simply applying an external magnetic field. Such a breakthrough allows for unprecedented control over the physical properties of materials, paving the way for new technological advances. However, progress has been sporadic due to the challenge of identifying effects that demonstrate the coupling between magnetism and electronic structure. The complexity of electronic structures often leads to unpredictable outcomes in the coupling mechanism with the magnetic state. The case of the antiferromagnet \nips\ illustrates this conundrum, where the observation of narrow emission peaks at \neel\ temperature has sparked long-standing debates about their magnetic origins~\cite{hwangbo_highly_2021,wang_spin-induced_2021,belvin_exciton-driven_2021,jana_magnon_2023}. Accurately determining the nature of the ground and excited states within the van der Waals magnet is crucial to unlocking the full potential of two-dimensional \vdW\ magnets.\\

Chromium Thiophosphate (\CPS) has emerged as a promising candidate for tuning the magnetic phase diagram across various thicknesses using an electrostatic gate~\cite{son_air-stable_2021,wu_gate-controlled_2023}. The properties that have drawn significant attention to this material are its air-stability~\cite{son_air-stable_2021} and the increased conduction bandwidth, especially compared to other comparable magnetic \vdW\ semiconductors. The latter feature allowed transport studies that revealed not only a new universal mechanism for magnetoconductance~\cite{wu_magnetism-induced_2023}, but this layered antiferromagnetic compound raised also a lot of interest because of strong proximity effects when brought in contact with other materials~\cite{wu_magnetotransport_2022} and unique magnon transport characteristics~\cite{de_wal_long-distance_2023}. The optical properties have also been recently revisited~\cite{diehl_crystal_1977,louisy_physical_1978,lee_structural_2017,gu_photoluminescent_2020,peng_controlling_2022,kim_photoluminescence_2022,susilo_band_2020,riesner_temperature_2022}, however, without leading to a definitive picture of the complete electronic structure of the material. For instance, the role of the magnetic transitions on the optical emission and absorption properties remains under debate~\cite{gu_photoluminescent_2020,riesner_temperature_2022}.\\

  \begin{figure}
 		\centering
 		\includegraphics[width=.95\linewidth]{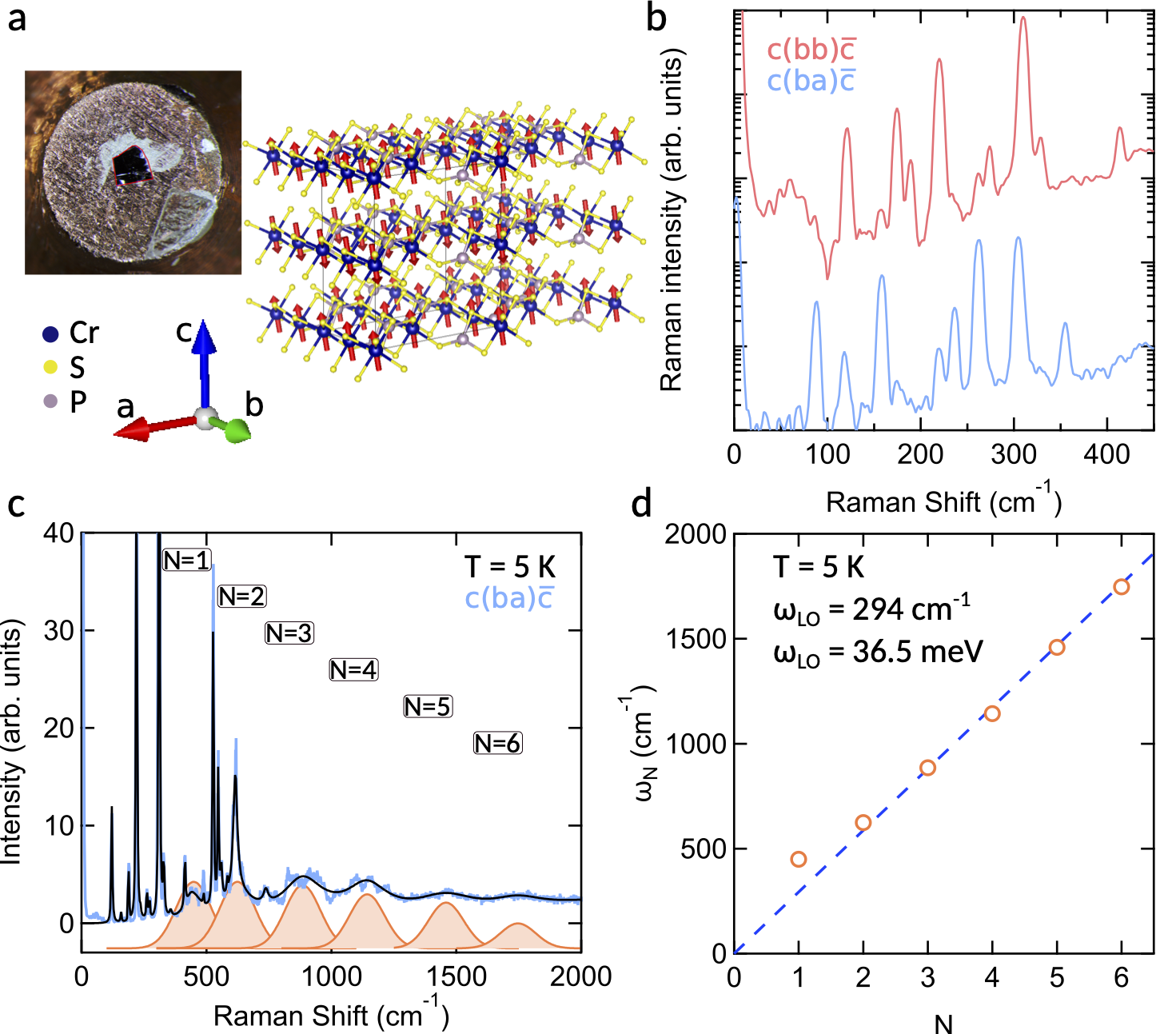}
 		\caption{ (\textsf{a}) Optical micrograph of a representative \CPS\ crystal used for this work mounted on the sample holder. On the right, the crystallogrphic structure if presented. (\textsf{b}) Polarization resolved Raman scattering spectra. Excitation of the laser beam is aligned with the b-axis and the analyzer aligned with the a- (blue) and b-axis (red). (\textsf{c}) Broad band spectrum up to 2000~\cm\ showing the multi-phonon character in \CPS. (\textsf{d}) Position of the harmonic peaks of the LO mode as function of mode number. The slopes yields a value of 36.5~meV (294~\cm) of the non-raman active LO mode.}
 	\label{fig:1}
 	\end{figure}
  
Here, we reveal a luminescent transition in chromium thiophosphate that emerges at higher energies compared to the lowest emission and absorption as the temperature approaches the \neel\ transition. Our spectroscopic analysis over a wide temperature and spectral range indicates that \CPS\ exhibits a complex spectrum, highlighting a strong interconnection between lattice, orbital, spin and charge degrees of freedom and optical properties. A key finding is a strong electron-phonon coupling, which favors the emergence of the luminescent transition at approximately 2~eV. The optical transitions are attributed to localized dd-transitions between the crystal field split states of the \CrPlus\ atoms, affected by distortion from the ideal octahedral environment near the \neel\ temperature. This asymmetric distortion along the b-axis suggests that \CPS\ is polar, hinting at the potential for multiferroicity in the material. Our findings not only establish a correlation between magnetism and the chromium ion environment evolution in \CPS, but also enhance our understanding of the electronic ground and excited state configurations of the material, underlining the complex inter-dependencies that govern its properties.\\

\begin{figure*}
 		\centering
 		\includegraphics[width=.85\linewidth]{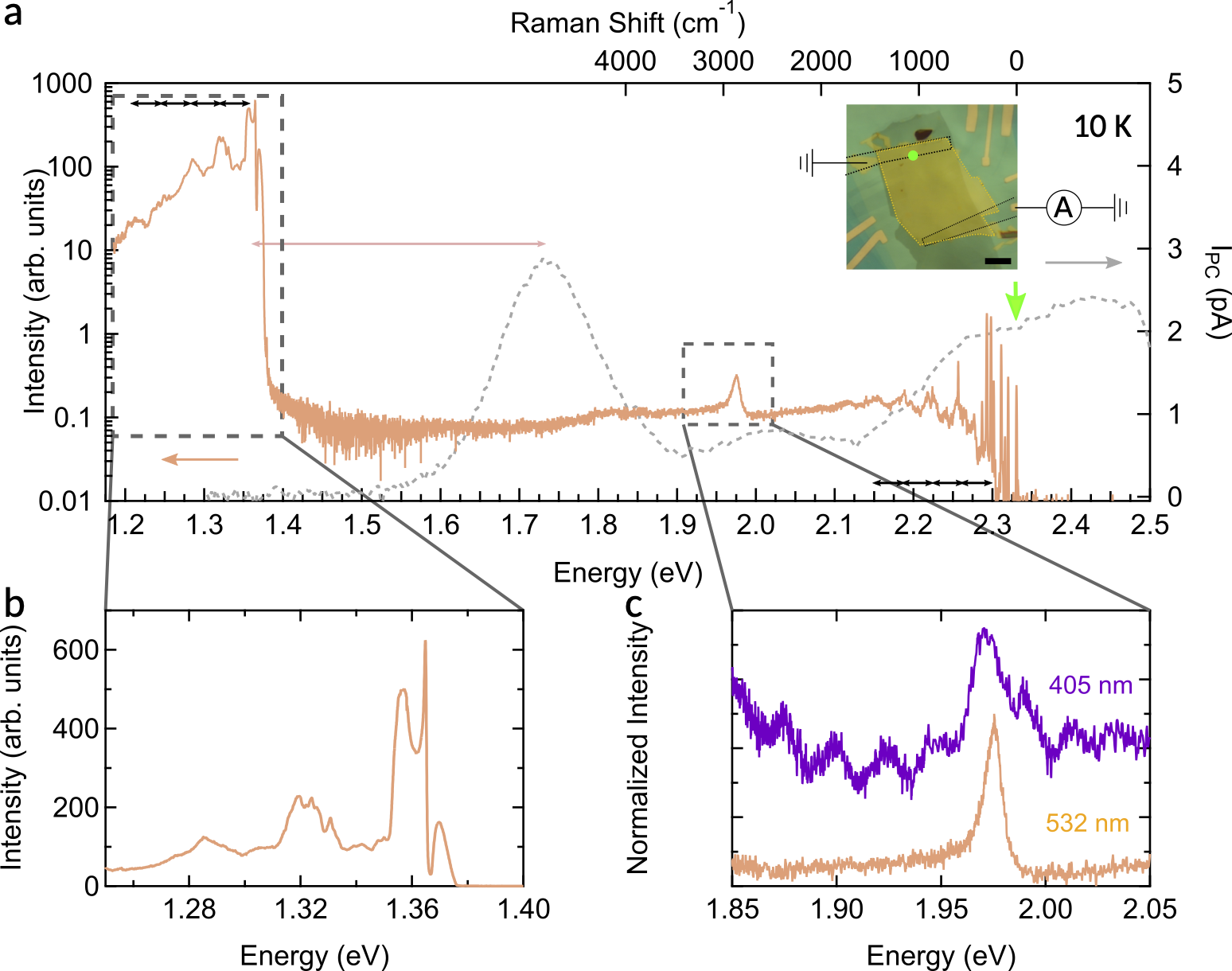}
 		\caption{ (\textsf{a}) Full emission spectrum (solid pink) of \CPS\ after excitation with a 2.33~eV laser (532~nm, green arrow). Raman lines are observed close to the excitation energy, while at lower energies the peaks can be attributed to luminescence processes. Dashed grey curve represents the photocurrent of the device shown in the inset, proportional to the absorption of the material. Black dashed line represents the thin graphite contacts attached to the sample and the green dot the position at which the spectrum was obtained. The Stokes shift --energy difference between the  absorption and the emission energy-- is about 350~meV and shown by the pink arrow. (\textsf{b}) Zoom in of the low energy photoluminescence where the main peak is centered around 1.36~eV. The lower energy peaks are spaced by about 35~meV (also indicated by the arrows in panel \textsf{a}) which is of the same period than the multi-phonon peaks observed in the Raman spectrum. (\textsf{c}) Comparison of the 2~eV emission spectrum after excitation with 2.33~eV laser (yellow) and 3.06~eV (405~nm) laser (plum colored curve).}
 	\label{fig:2}
		
 \end{figure*}
  
We start investigating the optical properties of \CPS\ bulk crystals down to low temperature with a detailed characterization by Raman spectroscopy. We present the crystal structure and an optical micrograph of a bulk crystal of \CPS\ in Figure~\figref{fig:1}{a}, known to be a layered antiferromagnetic (AFM) semiconductor with a \neel\ transition temperature at 38~K~\cite{diehl_crystal_1977,calder_magnetic_2020,peng_magnetic_2020,budko_magnetic_2021}. The Raman modes displayed in Figure~\figref{fig:1}{b} agree with the findings of previous reports that allow to determine the complete Raman tensor of the material (see the Methods section for details of the experimental setups used in this work)~\cite{kim_polarized_2021}. The polarization sensitive peaks enable us to ascertain the precise orientation of the crystallographic axes. The phonon dispersion of \CPS\ predicts modes extending up to 450~\cm~\cite{zhuang_density_2016}; however, the spectrum exhibit high-order modes and a periodic broad mode up to energies as high as $\approx$~2000~\cm, as shown in Figure~\figref{fig:1}{c}. By plotting the energy of each broad mode against its corresponding order, as depicted in Figure~\figref{fig:1}{d}, we determine the energy of the fundamental mode to be approximately 36.5~meV, typically associated to a longitudinal optical (LO) phonon. Similar findings have been reported in related materials like \CrIThree\ and transition-metal monochalcogenides~\cite{jin_observation_2020,klein_multiple-phonon-resonance_1969,leite_multiple-phonon_1969,merlin_multiphonon_1978,osterhoudt_charge_2018} and are the hallmark of the presence of strong electron-phonon coupling. The presence of large electron-phonon coupling is of crucial importance as vibronic states in \CPS\ are of primary importance in the interpretation of the optical data which will follow.\\


 \begin{figure*}
 		\centering
 		\includegraphics[width=.75\linewidth]{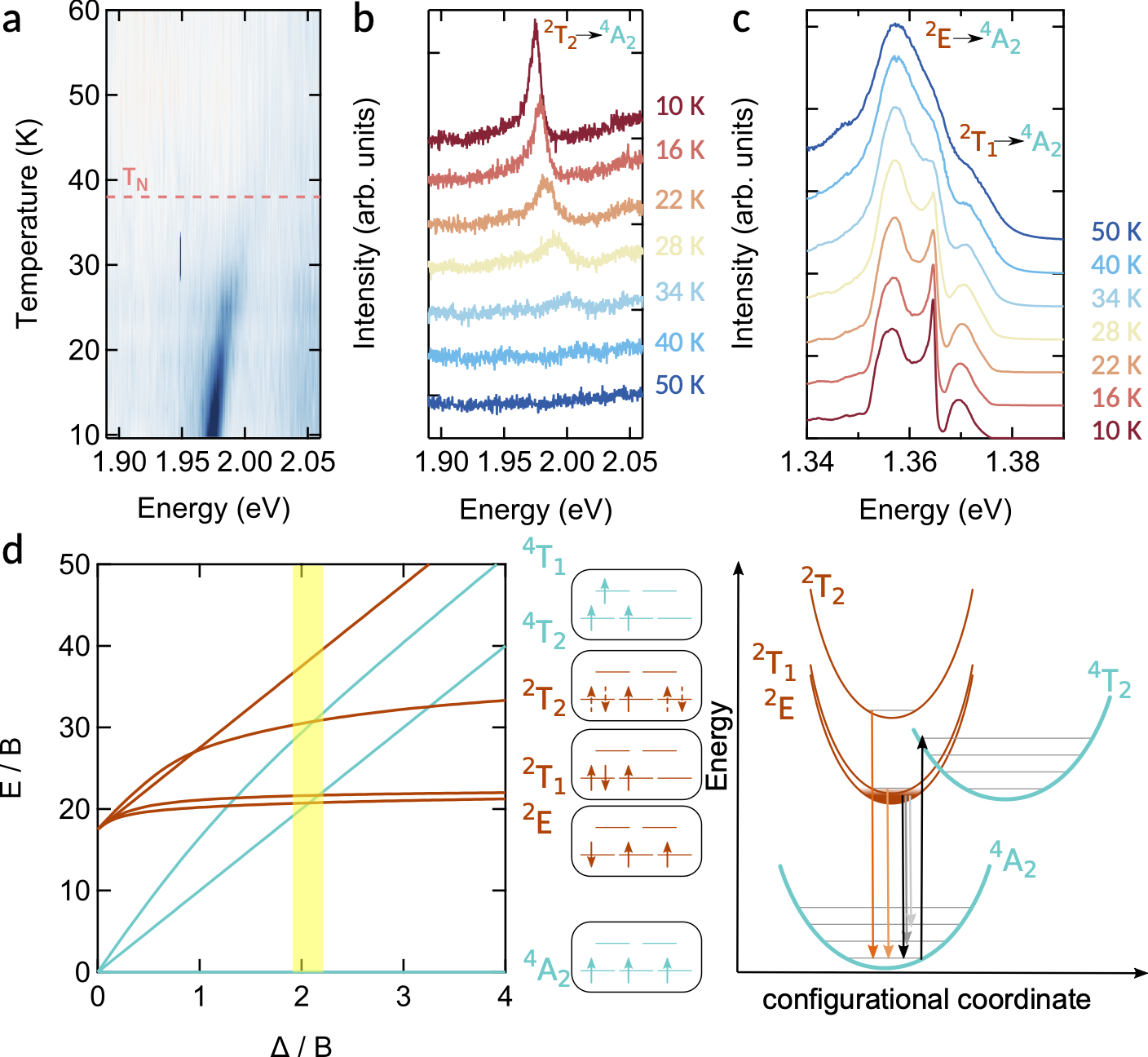}
 		\caption{ (\textsf{a}) False colorplot of the intensity of the 2~eV emission as a function of photon energy and temperature. The AFM transition temperation, \Tn~=~38~K, is indicated by the red dashed horizontal line. (\textsf{b}) Evolution of the spectrum at selected temperature indicated in the legend of the figure. (\textsf{c}) Evolution of the spectrum at selected temperature indicated in the legend of the panel for the low energy photoluminescence, showing the emergence of a Fano-like feature at 1.365~eV. (\textsf{d}) Tanabe-Sugano fan-chart (left) and the configurational orbital bands (center) and coordinate diagram (right) of the 3d \CrPlus.}
 	\label{fig:3}
 \end{figure*}
 
 
We now turn our attention to the broad band emission spectrum ranging from 1.2 to 2.2~eV after exiting the \CPS\ crystal with a 2.33~eV (532~nm) laser at the base temperature of our cryostat (10~K) (the solid yellow line in Fig.~\figref{fig:2}{a}). A photoluminescence peak near 2~eV immediately draws attention, yet, to fully capture its origin, we focus first on more discernible features of the spectrum. We identify the Stokes sharp Raman lines discussed earlier at slightly lower energies than the excitation line marked by the green arrow in Fig.~\figref{fig:2}{a}. In the lowest energy range of the spectrum we detect a pronounced photoluminescence signal centered around 1.35~eV, which is rich in spectroscopic features as seen in Figure~\figref{fig:2}{b}~\cite{gu_photoluminescent_2020,riesner_temperature_2022,kim_photoluminescence_2022}. We immediately note that the optical properties of \CPS\ diverge significantly from the ones of conventional semiconductors. As shown in Figure~\figref{fig:2}{a}, we observe a notable shift between the photoluminescence emission and the main absorption line, the latter obtained through photocurrent measurements performed on a 20~nm thick crystal with graphite contacts (see the inset of Figure~\figref{fig:2}{a}). A Stokes shift is commonly observed in this class of materials, however \CPS\ stands out as the value of about 350~meV found here is among the largest reported~\cite{larsen_photoconductivity_1972}. This finding points toward the fact that the local electrostatic environment is of importance for the electronic and optical properties of the material.\\
 
A close inspection of the emission peak centered around 1.35~eV confirms the conclusion and reveals that this photoluminescence displays an oscillating behavior with a period nearly identical to that observed in the Raman data, around 35~meV. This periodicity, coupled with the observed Stokes shift, leads to a Huang-Rhys factor of $S~=~5$, signifying very strong electron-phonon coupling within the material confirming the conclusion inferred earlier from the Raman spectroscopy. Additionally, on the high-energy shoulder of the main PL peak, a sharp Fano-like feature at 1.365~eV is detected, corroborating previous reports in this material~\cite{gu_photoluminescent_2020,riesner_temperature_2022}. However, as pointed out earlier, the luminescence peak at 1.35~eV intriguingly does not represent the only photoluminescence activity of the material.\\

 \begin{figure}
 		\centering
 		\includegraphics[width=.99\linewidth]{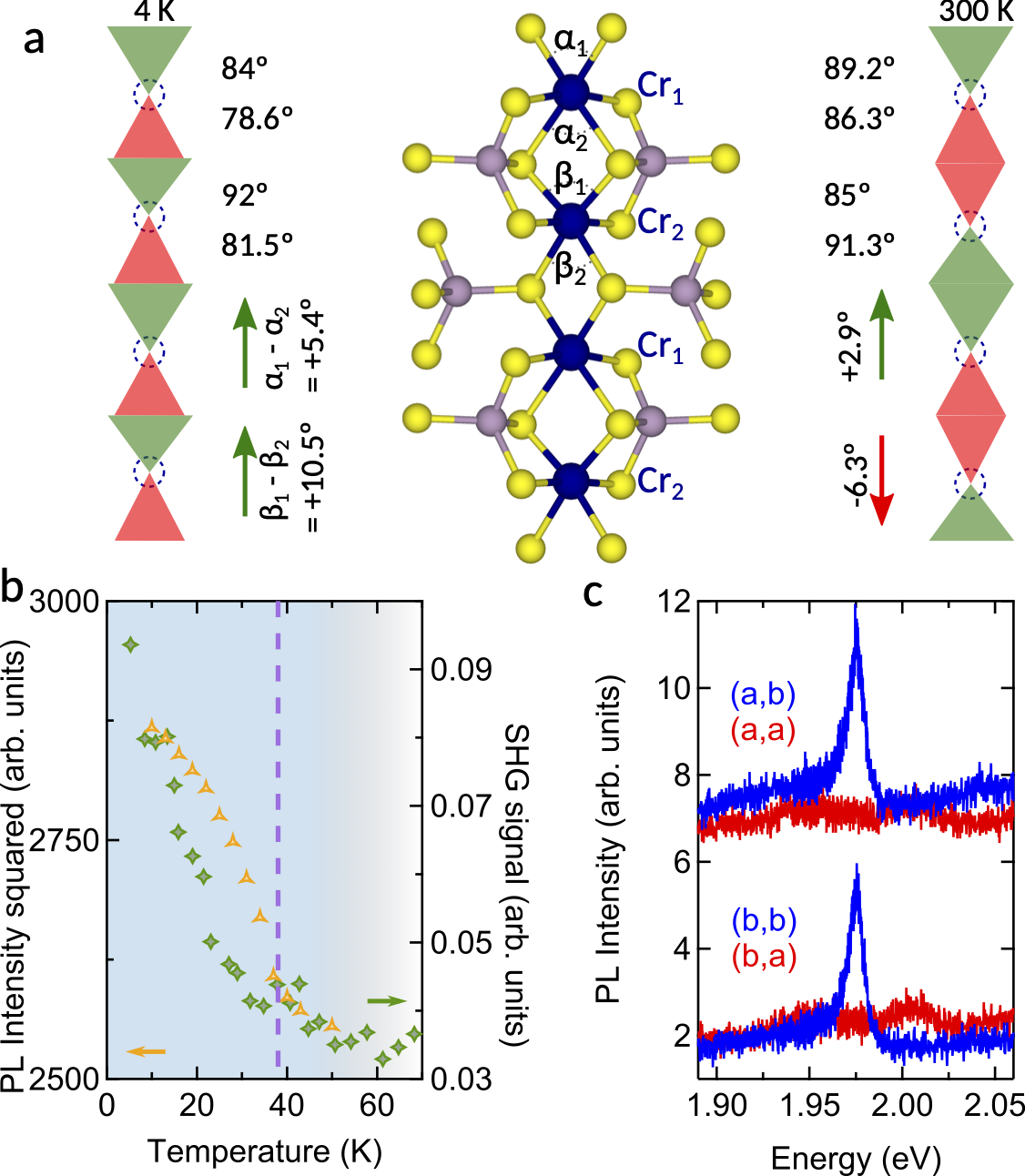}
 		\caption{ (\textsf{a}) Top-view of the crystal structure of \CPS\ showing the distortion angle between the Cr and S bonds in the basal plane at 4~K (left) and 300~K. The arrows show the polarity of the distortion which changes when going to low temperatures. (\textsf{b}) Temperature dependence of the squared intensity of the photoluminescence 2~eV peak (yellow, left axis) and the intensity of the SHG (green, right axis). We notice a nearly perfect ($\approx$~94~\%) correlation between these quantities. (\textsf{c}) Polarization resolved photoluminescence of \CPS\ showing extremely strong polarization of the transition along the b-axis regardless of the polarization of the incident laser.}
 	\label{fig:4}
 \end{figure}
 
In our low-temperature emission spectrum, we identify an additional peak, albeit weak, centered at 1.97~eV, as illustrated in Figure~\figref{fig:2}{a,c}. Remarkably, the energy of this emission peak is higher than the lowest absorption peak of the material, a phenomenon not often encountered encountered in semiconductor optics. Most likely it is the high position in energy that accounts for the low oscillator strength, which is 3 orders of magnitudes lower than the main photoluminescence peak and of the same intensity as the Raman signal. However, as shown in Figure~\figref{fig:2}{c}, upon varying the excitation wavelength of the laser (here from 2.33~eV to 3.06~eV) the energy of the emission line remains the same, indicating a luminescent nature rather than being the result of a Raman scattering process.\\

A question that immediately arises is about the physical mechanism behind these optical transitions and the impact of the magnetic properties of \CPS\ on its photoluminescence characteristics, especially concerning the newly observed 2~eV transition. Our subsequent analyses, illustrated in Figures~\figref{fig:3}{a} and \figref{fig:3}{b}, reveal that this 2~eV transition is indeed activated around the same temperature as the AFM phase transition occurring in the material. Interestingly, we note that the temperature dependence of the Fano-like resonance observed at 1.365~eV follows a similar trend (see Figure~\figref{fig:3}{c}), suggesting that both lines originate from a similar physical effect linked to the entrance into the AFM state.  The relationship between magnetism and the optical characteristics of \CPS\ is subject of debate~\cite{gu_photoluminescent_2020,riesner_temperature_2022}. Nevertheless, the significant Stokes shift and pronounced electron-phonon coupling, together with the observation of a large absorption threshold feature above 2~eV in absorption measurements, lead us to the conclusion that the optical transitions ranging between 1.3 and 2~eV originate from localized dd-transitions on the \CrPlus\ atoms. We stress that optical spectroscopy and tunneling spectroscopy provide complementary information on the electronic structure. Dd-excitations are collective modes which can be detected with optical techniques. In contrast, photo-emission and tunneling spectroscopy probe the single electron band structure.\\

Understanding the origin of the 2~eV transition necessitates to consider the energy diagram of the \CrPlus states in an octahedral environment depicted by the Tanabe-Sugano fan-chart shown in Figure~\figref{fig:3}{d} (alongside the configurational orbital bands and coordinate diagram) in the present the $d3$ configuration. At first glance, this interpretation seems contradicting recent reports of electron transport in \CPS~\cite{wu_gate-controlled_2023,wu_magnetism-induced_2023} which imply a significant Cr-S hybridization leading to a dispersive conduction bandwidth, although some of the valence bands are weakly dispersing~\cite{ohno_reflection_1989,zhuang_density_2016}. More important is that the dd-excitations are, by virtue of being many-body entangled states, rather robust in the presence of dispersion of the single electron bands.  As we will discuss later, the \CPS\ are not perfect octahedra, which causes a splitting of the multiplets shown in Figure~\ref{fig:3}. From the bandstructure calculations we can estimate these splittings to be of order of tens of meV. While being mindful of those refinements, Figure~\figref{fig:3}{d} provides an initial insight into the underlying mechanisms.\\

At room temperature, absorption and emission processes are  well described by spin-allowed transitions of electrons from the \QuAtwo\ ground state and the \QuTtwo. As a result of the on-site Coulomb and exchange interactions, the eigenstates of 3 electrons in the same 3d shell are in most cases not described by simply putting each of these electrons in an orbital state but require a linear combination of several Slater determinants (see supplementary information for details). Upon cooling down \CPS, electrons optically excited to the \QuTtwo\ state undergo intersystem crossing (ICS) with the \HaE\ state at about 150~K, shifting the broad luminescence  peak from 1.1~eV (\QuTtwo\ to \QuAtwo) to a central energy around 1.35~eV (\HaE\ to \QuAtwo)~\cite{kim_photoluminescence_2022}. Note that the orbital occupation of the ground state (\QuAtwo) and the excited \HaE\ is identical and their energy difference is due to the on-site exchange interaction. The oscillations on the lower energy shoulder stems from transition into higher lying vibrational states of the ground state (see the configurational coordinate diagram in Figure~\figref{fig:3}{d}). Prior estimations of crystal field splitting~\cite{riesner_temperature_2022}, represented by $\Delta/B$ (where $B$ is the Racah parameter indicating electron-electron interactions, including spin-orbit coupling and electron-electron repulsion), suggest a ratio of about 2, highlighted in Figure~\figref{fig:3}{d}  by the yellow shaded area. Following this estimate of the crystal field strength we find that the emission energy of the 2~eV peak almost perfectly matches the \HaTtwo\ energy level, while the Fano-like resonance originates from the \HaTone\ energy level into the ground state. However, as can be seen from the orbital configuration depicted in Figure~\figref{fig:3}{e}, transitions from the $^2$T states are spin-forbidden in the crystal field of an ideal octahedron, which indicates that around the AFM transition changes occur in the \CrPlus\ environment.\\

A detailed examination of the crystal field environment surrounding the \CrPlus\ ions presented in Figure~\figref{fig:4}{a} sheds light on a possible mechanism behind the activation of these transitions. In fact, each \CrPlus\ (S~=~3/2) ion is coordinated by six sulfur atoms in the form of a slightly distorted octahedron as illustrated by the difference in the S-Cr-S bond angle ($\alpha_1 - \alpha_2$ and $\beta_1 - \beta_2$ for two adjacent Cr$_1$ and Cr$_2$ sites, respectively)~\cite{ohno_reflection_1989}. Structural analyses~\cite{diehl_crystal_1977,budko_magnetic_2021} reveal a transition from anti-polar to polar alignment of the octahedra as temperatures drop, illustrated in Figure~\figref{fig:4}{a}.  This configurational change is signaled by an enhancement of the SHG intensity as the temperature approaches the \neel\ transition as shown in Figure~\figref{fig:4}{b}. A nearly perfect correlation is observed between the SHG signal and the square of the photoluminescence intensity of the 2~eV transition as a function of temperature. The SHG is therefore proportional to the square of the static dipole generated by the distortion. The fully linear polarized PL along the crystallographic b-axis, \ie\ parallel to the polar moment (see Figure~\figref{fig:4}{c}) is a clear indication that also the optical transition is proportional to the static dipole. The modified distortion of the ideal octahedral symmetry can lift the degeneracy of the wavefunctions of the excited state and activate transitions that are initially forbidden through mixing of the constituent wavefunctions of the multiplets (see Supplementary Information)~\cite{schmidt_exciton-magnon_2013}. Alternatively, the brightening of the 2~eV transition might be attributed to individual \CrPlus\ sites acting as quantum emitters, shifting from out-of-phase emission at higher temperatures, to in-phase at lower temperatures, following the polar-distortion long range arrangement. While further investigation is required to fully understand these phenomena, the observable changes in the polar environment, as indicated by SHG, are clearly linked to the activation of the \HaTone\ and \HaTtwo\ transitions.\\

The modification in the configuration of the distorted octahedral form is likely to induce an electrical polarization in the layers of \CPS. The increase of the SHG intensity corroborates this conclusion because it is particularly sensitive to inversion symmetry breaking, which for instance arises from electric fields~\cite{song_evidence_2022}. Note, that neither Raman spectroscopy nor neutron diffraction measurements indicate further symmetry lowering induced by the interlayer antiferromagnetic ordering of \CPS\ which excludes that the emerging SHG arises from a similar effect as reported for bilayer CrI$_3$~\cite{sun_giant_2019}. Recent estimates on the other hand have inferred a polarization value around 32~pC$\cdot$m$^{-2}$ at room temperature in the basal plane~\cite{neal_exploring_2021}. Following the trends observed in our SHG data and the analyses presented in Figure~\figref{fig:4}{a}, we expect an increase of this value at low temperature. While antiferromagnetism of the compound below 38~K is well established, the question of whether the system also exhibits ferroelectric (switchable polarization) or pyroelectric (non-switchable polarization) characteristics remains open. This potential multiferroicity in \CPS\ underscores the complex interplay between its structural, electronic, and magnetic properties, marking it as a material of considerable interest for future research.\\

In conclusion, we report the brightening of a  high-energy transition within the emission spectrum of the magnetic semiconductor \CPS, which emits at energies above the lowest absorption level and signals the appearance of magnetic order; a rare phenomenon  among magnetic materials. We attribute the brightening of the transition to a transformation within the distorted octahedral environment of the \CrPlus\ ions leading to a polar alignment and facilitating spin-forbidden dd-transitions within the 3d electron configuration. Complementary SHG measurements evidence that the distortion induces polarization within the basal planes of \CPS, particularly below the \neel\ transition temperature,  hinting to the existence of multiferroicity in the material. Our results underscore the tight link between lattice dynamics and electronic transition and allow an understanding of the structure of ground and excited state in the material above and below the \neel\ temperature. The disentanglement of relationship between electronic, crystallographic, and magnetic properties of the \CPS\ heralds new functional capabilities based on the compound for optoelectronic device applications.\\


\begin{acknowledgments}
The authors gratefully acknowledge fruitful discussions with Marco Gibertini, Thomas Olsen, and Sara López Paz. The data that support the findings of this study are openly available in the yareta portal of the University of Geneva (\href{https://yareta.unige.ch/home}{https://yareta.unige.ch/home}) as soon as possible.
\end{acknowledgments}

\bibliography{CPS4_NU}

\end{document}